\documentclass[preprint,amsmath,amssymb,aps,showkeys]{revtex4}
\usepackage[english]{babel}
\usepackage{graphicx}
\usepackage{graphics}
\usepackage{amsmath}
\usepackage{dcolumn}
\usepackage{amssymb}
\usepackage{bm}

\begin{document}
\title{Induced electric dipole moment coupling in Dirac equation}
\author{K. Bakke}
\email[E-mail address: ]{kbakke@fisica.ufpb.br}
\homepage[Orcid: ]{https://orcid.org/0000-0002-9038-863X}
\affiliation{Departamento de F\'isica, Universidade Federal da Para\'iba, Caixa Postal 5008, 58051-900, Jo\~ao Pessoa, PB, Brazil.}

\begin{abstract}

We introduce a nonminimal coupling into the Dirac equation as a model of describing the interaction of a Dirac neutral particle with an induced electric dipole moment with magnetic and electric fields. Then, we obtain a relativistic geometric quantum phase from the interaction of the induced electric dipole moment with electric and magnetic fields. Further, we include the permanent magnetic dipole moment of the Dirac neutral particle besides the induced electric dipole moment and obtain the relativistic geometric quantum phase.

\end{abstract}

\keywords{relativistic geometric quantum phase, induced electric dipole moment, Aharonov-Bohm effect, Aharonov-Casher effect, He-McKellar-Wilkens effect}

\maketitle

\section{Introduction}

Geometric quantum phases are a fascinating theme that can be found from studies of persistent currents \cite{by,tan,dantas2,spin1,spin2,spin3,spin4} to holonomic/geometric quantum computation \cite{qhal,qhal2,qhal3,qhal4,qhal5,qhal6,qhal7}. The most well known quantum effect related to geometric quantum phases is the Aharonov-Bohm effect \cite{ab}. Other quantum effects associated with geometric quantum phases are the scalar Aharonov-Bohm effect \cite{zei,pesk}, the dual Aharonov-Bohm effect \cite{dab,dab1}, the Aharonov-Casher effect \cite{ac} and the He-McKellar-Wilkens effect \cite{hmw,hmw2}. Inspired by these quantum effects, Wei {\it et al} \cite{wei} showed that a geometric quantum phase arises from the interaction of the induced electric dipole moment of a neutral particle with electric and magnetic fields. Later, Furtado {\it et al} \cite{lin} proposed an analogue of the Landau quantization for this system of a neutral particle based on the Wei {\it et al} proposal \cite{wei}. In another perspective, Audretsch {\it et al} \cite{bessel3} explored the attractive inverse-square potential which stems from the interaction of an electric field with the induced electric dipole moment of a neutral particle in search of bound states. From the studies of Refs. \cite{lin,bessel3}, neutral particles with induced electric dipole moment have been the theme of several works \cite{dantas,bf4,ob,br,br2,bf6,bf5}. In search of a relativistic description of the Wei {\it et al} system \cite{wei}, Refs. \cite{ind1,ind2} dealt with a scalar field and showed the Lagrangian for such interaction. However, a relativistic description of the Wei {\it et al} geometric quantum phase \cite{wei} based on a Dirac field has not been proposed.

In this work, we propose to introduce a nonminimal coupling into the Dirac equation with the aim of giving a relativistic description of a Dirac neutral particle with an induced electric dipole moment which interacts with electric and magnetic fields. We show that a relativistic geometric quantum phase can stem from the interaction of the induced electric dipole moment with electric and magnetic fields. Further, we consider a Dirac neutral particle with a permanent magnetic dipole moment besides the induced electric dipole moment, and then, obtain the relativistic Anandan quantum phase \cite{ana,anan1}.

The structure of this paper is: in section II, we introduce a nonminimal coupling into the Dirac equation in search of a relativistic description of a Dirac neutral particle with an induced electric dipole moment which interacts with electric and magnetic fields. Then, we discuss the appearance of a relativistic geometric quantum phase for a Dirac neutral particle with an induced electric dipole moment. We go further by discussing the relativistic geometric quantum phase for a Dirac neutral particle with a permanent magnetic dipole moment besides the induced electric dipole moment; in section III, we present our conclusions.

\section{Dirac neutral particle with an induced electric dipole moment}

Let us begin by defining the electromagnetic field tensor through its components as $F_{0i}=-E_{i}$ and $F_{ij}=\epsilon_{ijk}\,B^{k}$ ($\vec{E}$ is the electric field, while $\vec{B}$ is the magnetic field) and the metric $\eta^{\mu\nu}=\mathrm{diag}\left(-+++\right)$ \cite{carroll}. Besides, we shall work with $\hbar=1$ and $c=1$. According to Refs. \cite{ind1,ind2} we can define the moment tensor of atoms $K_{\mu\nu}$ from $F_{\mu\nu}$ by replacing $\vec{E}$ with $\vec{P}$ and $\vec{B}$ with $-\vec{M}$. In addition, by considering only the induced moments, we can write $\vec{P}=\alpha\vec{E}$ and $\vec{M}=\chi\,\vec{B}$, where $\alpha$ is the electric polarizability and $\chi$ is the magnetic susceptibility. Thereby, the components of the tensor $K_{\mu\nu}$ can be given by $K_{0i}=-\alpha\,E_{i}$ and $K_{ij}=-\chi\,\epsilon_{ijk}\,B^{k}$. In this work, our proposal is to introduce a nonminimal coupling into the Dirac equation as follows:
\begin{eqnarray}
i\,\gamma^{\mu}\,\partial_{\mu}\rightarrow i\,\gamma^{\mu}\,\partial_{\mu}-\frac{1}{4}\,\eta^{\alpha\beta}\,K_{\mu\alpha}\,F_{\beta\nu}\,\gamma^{\mu}\,\gamma^{\nu}.
\label{1.1}
\end{eqnarray}
Note that we have neglected in Eq. (\ref{1.1}) the intrinsic moment (proportional to the spin of the neutral particle \cite{ind1,ind2}). We wish to focus only on the induced moments, which are proportional to the applied electric and magnetic fields. The $\gamma^{\mu}$ matrices are defined as \cite{greiner}
\begin{eqnarray}
\gamma^{0}=\hat{\beta}=\left(
\begin{array}{cc}
I & 0\\
0 & -I\\
\end{array}\right);\,\,\,\,
\gamma^{i}=\hat{\beta}\hat{\alpha}^{i}=\left(
\begin{array}{cc}
0 & \sigma^{i}\\
-\sigma^{i} & 0\\
\end{array}\right);\,\,\,
\Sigma^{i}=\left(
\begin{array}{cc}
\sigma^{i} & 0\\
0 & \sigma^{i}\\
\end{array}\right),
\label{1.2}
\end{eqnarray}
where $\sigma^{i}$ are the Pauli matrices, $I$ is the $2\times2$ identity matrix and $\vec{\Sigma}$ is the spin vector. Thereby, after some calculations, the Dirac equation becomes
\begin{eqnarray}
i\frac{\partial\Psi}{\partial t}&=&m\,\hat{\beta}\,\Psi+\vec{\alpha}\cdot\hat{p}\Psi+\frac{1}{4}\hat{\beta}\,\vec{\alpha}\cdot\left[\alpha\vec{E}\times\vec{B}\right]\Psi-\frac{1}{2}\,\alpha\,E^{2}\,\hat{\beta}\,\Psi\nonumber\\
&+&\frac{1}{4}\hat{\beta}\,\vec{\alpha}\cdot\left[\chi\vec{E}\times\vec{B}\right]\Psi-\frac{1}{2}\,\chi\,B^{2}\,\hat{\beta}\,\Psi.
\label{1.3}
\end{eqnarray}

With the purpose of studying geometric quantum phases for a Dirac neutral particle with an induced electric dipole moment, let us take $\chi=0$ as in Refs. \cite{ind1,ind2}. Furthermore, let us consider the field configuration proposed by Wei {\it el al} \cite{wei,wei2}:
\begin{eqnarray}
\vec{E}=\frac{\lambda}{r}\,\mathbf{r};\,\,\,\,\vec{B}=B_{0}\,\mathbf{z}.
\label{1.4}
\end{eqnarray}
where $r$ is the radial coordinate and $B_{0}>0$ is a constant. The parameter $\lambda$, in turn, can be a constant associated with a linear distribution of electric charges along the $z$-axis \cite{wei}, or $\lambda=\frac{\rho\,R^{2}_{0}}{2}$ is a constant related to the uniform volume charge density $\rho$ inside the cylinder of radius $R_{0}$ \cite{br}. The vectors $\mathbf{r}$ and $\mathbf{z}$ are unit vectors in the radial and $z$ directions, respectively \footnote{ Note that, by working with the cylindrical symmetry, the term $i\gamma^{\mu}\,\partial_{\mu}$ in the Dirac equation (\ref{1.3}) must be replaced with $i\,\gamma^{\mu}\,\partial_{\mu}\rightarrow i\,\gamma^{\mu}\,\frac{1}{h_{\mu}}\,\partial_{\mu}+\frac{i}{2}\sum_{k=1}^{3}\gamma^{k}\left[\frac{1}{h_{k}}\,\partial_{k}\ln\left(\frac{h_{1}\,h_{2}\,h_{3}}{h_{k}}\right)\right]$, where $h_{\mu}$ are the scale factors of this coordinate system \cite{schu}. With the cylindrical symmetry, we have $h_{0}=h_{t}=1$, $h_{1}=h_{r}=1$, $h_{2}=h_{\varphi}=r$ and $h_{3}=h_{z}=1$.}. Hence, the Dirac equation for a neutral particle with an induced electric dipole moment is given by
\begin{eqnarray}
i\,\frac{\partial\Psi}{\partial t}&=&m\hat{\beta}\,\Psi-i\,\hat{\alpha}^{1}\left(\frac{\partial}{\partial r}+\frac{1}{2r}\right)\Psi-i\,\frac{\hat{\alpha}^{2}}{r}\frac{\partial\Psi}{\partial\varphi}-i\,\hat{\alpha}^{3}\frac{\partial\Psi}{\partial z}\nonumber\\
&+&\frac{1}{4}\hat{\beta}\,\vec{\alpha}\cdot\left(\alpha\vec{E}\times\vec{B}\right)\Psi-\frac{1}{2}\,\alpha\,E^{2}\,\hat{\beta}\,\Psi.
\label{1.5}
\end{eqnarray}

Since the fields given in Eq. (\ref{1.4}) are independents of the $z$-coordinate, thus, a particular solution to Eq. (\ref{1.5}) is given in the form: $\Psi\left(t,\,r,\,\varphi,\,z\right)=e^{i\,p_{z}\,z}\,\psi\left(t,\,r,\,\varphi\right)$, where $p_{z}$ is a constant. Then, we can reduce the system to a planar system by taking $p_{z}=0$, without loss of generality. Furthermore, when the neutral particle moves in a circular path, a geometric quantum phase arises from the interaction of the induced electric dipole moment with the fields given in Eq. (\ref{1.4}). By applying the Dirac phase factor method \cite{dirac1,dirac2}, we can write the wave function in the form: $\psi\left(t,\,r,\,\varphi\right)=\hat{\mathcal{P}}\,e^{i\Phi_{\mathrm{R}}}\,\psi_{0}\left(t,\,r,\,\varphi\right)$, where $\hat{\mathcal{P}}$ denotes the path ordering operator and $\psi_{0}\left(t,\,r,\,\varphi\right)$ is the solution to the following equation \footnote{With $\psi=e^{i\Phi_{\mathrm{R}}}\,\psi_{0}$, we have
$-i\,\frac{\hat{\alpha}^{2}}{r}\frac{\partial\psi}{\partial\varphi}=-i\,\frac{\hat{\alpha}^{2}}{r}\frac{\partial}{\partial\varphi}\left[e^{i\Phi_{\mathrm{R}}}\,\psi_{0}\right]=-\frac{\hat{\beta}}{4}\,\vec{\alpha}\cdot\left(\alpha\vec{E}\times\vec{B}\right)\,e^{i\Phi_{\mathrm{R}}}\,\psi_{0}-i\,\frac{\hat{\alpha}^{2}}{r}\,e^{i\Phi_{\mathrm{R}}}\,\frac{\partial\psi_{0}}{\partial\varphi}$. After substituting these terms into Eq. (\ref{1.5}), we achieve Eq. (\ref{1.6}).}:
\begin{eqnarray}
i\,\frac{\partial\psi_{0}}{\partial t}=m\hat{\beta}\,\psi_{0}-i\,\hat{\alpha}^{1}\left(\frac{\partial}{\partial r}+\frac{1}{2r}\right)\psi_{0}-i\,\frac{\hat{\alpha}^{2}}{r}\frac{\partial\psi_{0}}{\partial\varphi}-\frac{1}{2}\,\alpha\,E^{2}\,\hat{\beta}\,\psi_{0}.
\label{1.6}
\end{eqnarray}    

The relativistic geometric quantum phase $\Phi$, in turn, is given by 
\begin{eqnarray}
\Phi_{\mathrm{R}}=\frac{\hat{\beta}}{4}\oint\left(\alpha\vec{E}\times\vec{B}\right)\cdot d\vec{r}=-\frac{\pi}{2}\,\alpha\,\lambda\,B_{0}\,\hat{\beta}.
\label{1.7}
\end{eqnarray}

Hence, the relativistic geometric quantum phase (\ref{1.7}) is obtained without the adiabatic approximation, which corresponds to an Aharonov-Anandan quantum phase \cite{anan}. Moreover, it is a non-dispersive geometric quantum phase \cite{disp1,disp2,disp3} because it does not depend on the velocity of the particle. The geometric quantum phase (\ref{1.7}) can be viewed as the relativistic analogue of the geometric quantum phase obtained by Wei {\it et al} \cite{wei} for a Dirac neutral particle.  Note that, with the field configuration (\ref{1.4}), the term proportional to $E^{2}$ is a local term and does not contribute to the geometric quantum phase \cite{ana,anan1,lin2}. Actually, it plays the role of an attractive $r^{-2}$ potential \cite{bessel3,br2,bf6}.

Let us explore another perspective by taking into account the intrinsic moment, which are proportional to the spin of the neutral particle \cite{ind1,ind2}. In the case of a neutral particle with a permanent magnetic dipole moment in addition to the induced electric dipole moment (described by Eq. (\ref{1.1})), the covariant form of Dirac equation is
\begin{eqnarray}
m\Psi=i\,\gamma^{\mu}\,\partial_{\mu}\Psi-\frac{1}{4}\,\eta^{\alpha\beta}\,K_{\eta\alpha}\,F_{\beta\nu}\,\gamma^{\eta}\,\gamma^{\nu}\,\Psi+\frac{\mu}{2}\,\Sigma^{\beta\nu}\,F_{\beta\nu}\,\Psi,
\label{2.1}
\end{eqnarray}
where $\mu$ is the permanent magnetic dipole moment of the neutral particle and $\Sigma^{\beta\nu}=\frac{i}{2}\left[\gamma^{\beta},\,\gamma^{\nu}\right]$ \cite{bf,ana,anan1}. By taking $\chi=0$, we can write Eq. (\ref{2.1}) as follows:
\begin{eqnarray}
i\frac{\partial\Psi}{\partial t}&=&m\,\hat{\beta}\,\Psi+\vec{\alpha}\cdot\hat{p}\Psi+\frac{1}{4}\hat{\beta}\,\vec{\alpha}\cdot\left[\alpha\vec{E}\times\vec{B}\right]\Psi-\frac{1}{2}\,\alpha\,E^{2}\,\hat{\beta}\,\Psi\nonumber\\
&+&i\,\mu\,\hat{\beta}\,\vec{\alpha}\cdot\vec{E}\,\Psi-\mu\,\hat{\beta}\,\vec{\Sigma}\cdot\vec{B}\,\Psi.
\label{2.2}
\end{eqnarray}

In this case, the relativistic Anandan quantum phase \cite{ana,anan1,bf} is given by 
\begin{eqnarray}
\Phi_{\mathrm{A}}=\hat{\beta}\,\oint\left[\mu\,\,\vec{\Sigma}\times\vec{E}+\frac{1}{4}\alpha\vec{E}\times\vec{B}\right]\cdot d\vec{r}+\hat{\beta}\,\int_{0}^{\tau}\mu\,\vec{\Sigma}\cdot\vec{B}\,dt.
\label{2.3}
\end{eqnarray}
where $\tau$ corresponds to the time of interaction of the magnetic field with the magnetic dipole moment \cite{bf3}. By using the field configuration (\ref{1.4}) and with the magnetic dipole moment of the neutral particle aligned to the $z$-direction \cite{ac}, we obtain the relativistic geometric quantum phase:
\begin{eqnarray}
\Phi_{\mathrm{A}}=2\pi\,\mu\,\lambda\,\hat{\beta}\,\Sigma^{3}-\frac{\pi}{2}\,\alpha\,\lambda\,B_{0}\,\hat{\beta}+\mu\,B_{0}\,\tau\,\Sigma^{3}.
\label{2.4}
\end{eqnarray}
Note that the first term of Eq. (\ref{2.4}) corresponds to the relativistic analogue of the Aharonov-Casher effect \cite{ac}, while the last term of Eq. (\ref{2.4}) yields the scalar Aharonov-Bohm effect for neutral particles \cite{zei,bf3}. Therefore, from Eqs. (\ref{2.3}) and $(\ref{2.4})$, we observe that the relativistic Anandan quantum phase acquires a new contribution given by the interaction of the induced electric dipole moment of the Dirac neutral particle with electric and magnetic fields.

Finally, it is worth observing the Dirac equation (\ref{1.3}) when $\alpha=0$ and $\chi\neq0$. It yields the Dirac equation:  
\begin{eqnarray}
i\frac{\partial\Psi}{\partial t}&=&m\,\hat{\beta}\,\Psi-i\,\hat{\alpha}^{1}\left(\frac{\partial}{\partial r}+\frac{1}{2r}\right)\Psi-i\,\frac{\hat{\alpha}^{2}}{r}\frac{\partial\Psi}{\partial\varphi}-i\,\hat{\alpha}^{3}\frac{\partial\Psi}{\partial z}\nonumber\\
&+&\frac{1}{4}\hat{\beta}\,\vec{\alpha}\cdot\left(\vec{E}\times\chi\vec{B}\right)\Psi-\frac{1}{2}\,\chi\,B^{2}\,\hat{\beta}\,\Psi.
\label{3.1}
\end{eqnarray}

By using the field configuration given in Eq. (\ref{1.4}) and by following the steps from Eq. (\ref{1.4}) to Eq. (\ref{1.6}), we obtain the relativistic geometric quantum phase:
\begin{eqnarray}
\Phi_{\mathrm{R}}'=\frac{\hat{\beta}}{4}\oint\left(\vec{E}\times\chi\vec{B}\right)\cdot d\vec{r}=-\frac{\pi}{2}\,\chi\,\lambda\,B_{0}\,\hat{\beta}.
\label{3.2}
\end{eqnarray}

Therefore, Eq. (\ref{3.2}) shows the appearance of a relativistic geometric quantum phase for a Dirac neutral particle that interacts with crossed electric and magnetic fields when there is a non-null magnetic susceptibility $\chi$.

\section{Conclusions}

We have proposed the introduction of a nonminimal coupling into the Dirac equation with the aim of describing the interaction of the induced electric dipole moment with electric and magnetic fields from the Dirac equation. We have shown that a relativistic analogue of the Wei {\it et al} geometric quantum phase \cite{wei} can be achieved in this relativistic scenario. Besides, we have analysed the role of the term proportional to $E^{2}$ with the field configuration of Ref. \cite{wei}. It yields an attractive inverse-square potential. We have gone further by considering a neutral particle with a permanent magnetic dipole moment in addition to the induced electric dipole moment. We have seen that the relativistic Anandan quantum phase acquires a new contribution given by the interaction of the induced electric dipole moment of the Dirac neutral particle with electric and magnetic fields.

Finally, we have considered a scenario in which the induced electric dipole moment is null ($\alpha=0$) and magnetic susceptibility $\chi$ is non-null. With the field configuration of Ref. \cite{wei}, we have shown that the wave function of the Dirac neutral particle also acquires a relativistic geometric quantum phase determined by the magnetic susceptibility $\chi$ and the crossed electric and magnetic fields.

\acknowledgments{The author would like to thank CNPq for financial support.}

\section*{Data accessibility statement}

This work does not have any experimental data.

\section*{Competing interests}

We have no competing interests.

\section*{Ethics statement}

This research poses no ethical considerations.

\end{document}